\documentclass[11pt,a4paper]{article}

\RequirePackage[l2tabu, orthodox]{nag}

\usepackage{jheppub}
\usepackage{amsthm,amssymb,amsmath,epic,float}
\usepackage{rotating,epsfig,indentfirst,array,varioref}
\usepackage{appendix,tikz,mathtools}
\usepackage{setspace}
\usepackage{tikz}
\usetikzlibrary{decorations.pathreplacing}
\usetikzlibrary{calc}
\usepackage{enumitem}
\usepackage{hyphenat}
\usetikzlibrary{positioning}
\usepackage{graphicx}  
\usepackage{adjustbox}

\usepackage{arydshln}

\def\bbt{\bibitem}
\def\be{\begin{equation}}
\def\en{\end{equation}}
\def\ber{\begin{eqnarray}}
\def\enr{\end{eqnarray}}
\def\nmb{ \nonumber\\}

\def\rbr{\rbrack}
\def\lbr{\lbrack}
\def\rbrc{\rbrace}
\def\lbrc{\lbrace}
\def\ov{\over }
\def\tld{\tilde}

\def\sgm{\sigma}

\def\bet{\beta}

\def\im{\imath}

\def\dlt{\delta}

\def\vw {\vec{w}}
\def\vl{\vec{l}}
\def\vq{\vec{q}}
\def\vt{\vec{t}}

\def\vbt{\vec{\beta}}

\def\vk {\vec{k}}

\usepackage{jheppub}
\makeatletter
\def\@fpheader{\vspace{-.1cm}}
\makeatother
                                
\begin{document}

\title{\boldmath  Spectral flow construction of mirror pairs of CY orbifolds.}

\author [a]{Sergej Parkhomenko}

\affiliation[a]{Landau Institute for Theoretical Physics, 142432 Chernogolovka, Russia}


\abstract{We consider the CY orbifolds connected with the products of $N=(2,2)$ supersymmetric minimal models. We use the spectral flow construction of states,
as well as its mirror one, to explicitly show that these CY orbifolds are canonically combined into mirror pairs of isomorphic models, associated to the mutually dual pairs of admissible groups of Berglund-Hubsh-Krawitz.}



\maketitle

\section{Introduction}
\label{sec:intro}

As is known ten-dimensional superstring theory is a candidate for unification of the Standard Model and quantum gravity.
To get a four-dimensional theory with Space-Time supersymmetry
(which is necessary for phenomenological reasons),  we must compactify the 6 of 10 dimensions on the Calabi-Yau manifold, as shown by Candelas et al. \cite{CHSW}.
Another equivalent approach is the compactification of 6 dimensions into $N=(2,2)$ Superconformal field theory with the central charge $c=9$,  as shown by D. Gepner \cite{Gep}.

An important property of CY manifolds is the Mirror Symmetry, which reflects conjectured isomorphism of the $\sgm$-models for each mirror pair of CY manifolds. This symmetry was predicted by D. Gepner \cite{DG},
\cite{COGP} and discovered in \cite{CLS}, \cite{GrPl}, (see also \cite{LS}).

In paper \cite{GrPl} a construction was developed to show that partition functions of the orbifolds of Fermat-type CY manifolds coincide for each mirror pair. Their idea was based on the observation that orbifold of Minimal model $M_{k}$ over the group $\mathbb{Z}_{k+2}$ gives an equivalent, mirror reflected model. The more refined approach to the mirror symmetry was proposed by Borisov in \cite{B}. The author constructed certain vertex algebra for a CY hypersurface in toric manifold, which was endowed with an $N=2$ Virasoro superalgebra action and then proved that vertex algebra of mirror CY hypersurface is isomorphic to the original one. This paper was an important step towards rigorous construction of $N=(2,2)$ superconformal model of a CY manifold and proving the isomorphism for the models coming from mirror pairs of CY manifolds.

In this note, we consider orbifolds of CY hypersurfaces of Fermat-type given in projective varieties and show explicitly that the corresponding $N=(2,2)$ superconformal models are isomorphic for each mirror pair of such orbifolds. This note can also be considered as a natural continuation of the papers \cite{BP},\cite{BBP}, where the explicit construction of states of orbifolds of Fermat-type CY manifolds was developed. There we have used
the connection of CY orbifold models with the class of exactly solvable minimal $N=(2,2)$ superconformal models to explicitly construct a complete set of fields in the models
using spectral flow transformation 
and the requirement of mutual locality of fields. Here we present an equivalent but mirror spectral flow construction of states and show that it is naturally connected to the mirror orbifold model. More precisely, we show that the $N=(2,2)$ superconformal model associated with a CY orbifold with an admissible group $G_{adm}$ can also be treated as the $N=(2,2)$ superconformal model of a mirror CY orbifold with the Berglund-Hubsch-Kravitz dual group $G^{*}_{adm}$ if we apply the mirror spectral flow construction of states instead of the original one.

The plan of the paper is as follows. In Section~\ref{sec:2} we introduce the notations and briefly discuss the spectral flow construction of states in the Fermat-type CY orbifolds developed in \cite{BP}, \cite{BBP}. In Section~\ref{sec:3} we introduce the mirror spectral flow construction of states of these orbifolds. Then we prove that this mirror spectral flow construction allows us to consider the original $N=(2,2)$ superconformal CY orbifold model with admissible group $G_{adm}$ as a mirror CY orbifold constructed over the Berglund-Hubsh-Krawitz dual group $G^{*}_{adm}$. 

\section{The construction of orbifold states using spectral flow and mutual locality.}
\label{sec:2}

Here, we briefly discuss the explicit construction of states in the Fermat-type CY orbifolds developed in \cite{BP}, \cite{BBP}.

\paragraph{The composite of $N=(2,2)$ Minimal Models.}

We consider $\sgm$-models on the orbifolds of CY manifolds, which can be determined in the weighted projective space
$\mathbb{P}_{n_{1},...,n_{5}}$ in terms of zeros of the quasi-homogeneous polynomials of Fermat type
\ber
W(x_{i}):=\sum_{i=1}^{5}x_{i}^{k_{i}+2}, and \ W(\lambda^{n_i}x_{i})=\lambda^d W(x_{i}),
\ \text{where} \ d= \sum_{i} n_i.
\enr
According to Gepner \cite{Gep}, these kind of models are equivalent to
the orbifolds of the composite models constructed from the tensor products 
\ber
M_{\vk}=\prod_{i=1}^{5}M_{k_{i}}
\nmb
\enr 
of five $N=(2,2)$ superconformal minimal models $M_{k_{i}}$ with the total central charge $9$. 

A minimal superconformal model $M_{k}$ with $N=(2,2)$ symmetry, is labeled by positive integer $k$, which defines the central charge as
\ber
c={3k\ov k+2}.
\nmb
\enr 

 The commutation relations for the generators $L_n, J_{n}, G^{\pm}_{r}$ 
of $N=2$ Super-Virasoro algebra of symmetries of the model have the form
\ber
\begin{aligned}
&\lbr  L_{n},L_{m}\rbr=(n-m)L_{n+m}+{c\ov 12}(n^{3}-n)\dlt_{n+m,0},\\
&\lbr J_{n},J_{m}\rbr={c\ov 3}n\dlt_{n+m,0},\\
&\lbr L_{n},J_{m}\rbr=-mJ_{n+m},\\
&\lbrc G^{+}_{r},G^{-}_{s}\rbrc=L_{r+s}+{r-s\ov 2}J_{r+s}+{c\ov 6}(r^{2}-{1\ov 4})\dlt_{r+s,0},\\
&\lbr L_{n},G^{\pm}_{r}\rbr=({n\ov 2}-r)G^{\pm}_{r+n},\\
&\lbr J_{n},G^{\pm}_{r}\rbr=\pm G^{\pm}_{r+n}.
\label{2.N2Vir}
\end{aligned}
\enr

The space of fields in $NS$ and $R$ sectors can be decomposed into the sum of products of minimal representations
\ber
{\cal{H}}^{NS}_{l,q}, \ {\cal{H}}^{R}_{l,q}, \
l=0,1,...,k, \ q=-l,-l+2,...,l.
\label{2.MinRep}
\enr 
The conformal dimensions and charges of the corresponding primary 
states $\Phi^{NS}_{l,q}$ in $NS$ sector are
\ber
\Delta_{l,q}=\frac{l(l+2)-q^{2}}{4(k+2)}, \ \ Q_{l,q}=\frac{q}{k+2}.
\label{2.DeltQ}
\enr
In $R$ sector the dimensions and charges of primary states $\Phi^{R}_{l,q}$ are given by
\ber
\Delta^{R}_{l,q}=\frac{l(l+2)-(q-1)^{2}}{4(k+2)}+\frac{1}{8},\quad
\bar{Q}^{R}_{\l,q}=Q^{R}_{l,q}=\frac{q-1}{k+2}+\frac{1}{2}.
\label{2.DeltQR}
\enr
In general, the set of (quasi-) local fields of the minimal models have an $A-D-E$ classification, but in this paper we consider only $A$ series.  Then the primary fields of the model are given by diagonal pairing of holomorphic and anti-holomorphic factors:
\ber
&\Psi^{NS}_{l,q}(z,\bar{z})=\Phi^{NS}_{l,q}(z)\bar{\Phi}^{NS}_{l,q}(\bar{z}),
\\
&\Psi^{R}_{l,q}(z,\bar{z})=\Phi^{R}_{l,q}(z)\bar{\Phi}^{R}_{l,q}(\bar{z}),
\\
& \text{where}\quad l=0,...,k, \quad q=-l,-l+2,...,l.
\label{2.PrimNS}
\enr
and where the factors $\Phi^{NS}_{l,q}(z)$, $\bar{\Phi}^{NS}_{l,q}(\bar{z})$, $\Phi^{R}_{l,q}(z)$, $\bar{\Phi}^{R}_{l,q}(\bar{z})$ are the primary states of the minimal representations (\ref{2.MinRep})
so that their conformal dimensions and $U(1)$ charges  are given by (\ref{2.DeltQ}), (\ref{2.DeltQR}).

Recall that the chiral primary states \cite{LVW} appear when $q=l$:
\ber
\Phi^{c}_{l}\equiv \Phi^{NS}_{l,l},
\label{2.CPrim}
\enr
while the anti-chiral primary states \cite{LVW} appear when $q=-l$:
\ber
\Phi^{a}_{l}\equiv \Phi^{NS}_{l,-l}.
\label{2.APrim}
\enr

The conformal symmetry of the composite model $M_{\vk}$ is the Virasoro $N=(2,2)$   superalgebra, defined as the diagonal subalgebra in the tensor product of 5 minimal models as follows
\ber
L_{tot,n}=\sum_{i}L_{(i),n}, \ J_{tot,n}=\sum_{i}J_{(i),n},
\
G^{\pm}_{tot,r}=\sum_{i}G^{\pm}_{(i),r}.
\label{2.DiagVir}
\enr

The action of this algebra is correctly defined only on the product of $NS$ representations  or on the product of  $R$  representations of minimal models $M_{k_{i}}$.
Therefore, we can form $NS$ or $R$ primary fields  in the product model $M_{\vk}$ by taking only products of primary fields from each minimal model $M_{k_{i}}$  belonging to  the same ($ NS$ or $R$) sectors:

\ber
&\Psi^{NS}_{\vec{l},\vec{q}}(z,\bar{z})=\prod_{i}\Psi^{NS}_{l_{i},q_{i}}(z,\bar{z}),
\\
&\Psi^{R}_{\vl,\vq}(z,\bar{z})=\prod_{i}\Psi^{R}_{l_{i},q_{i}}(z,\bar{z}).
\label{2.Prim}
\enr

The descendant fields are generated from these primary fields  by the creation generators of the $N=2$ Virasoro superalgebras of the $M_{k_{i}}$ models. 

\vskip 10pt
\paragraph{The admissible group.}
To define the admissible group \cite{BH}, \cite{Kra} giving the orbifold model we recall that 
the composition of the $M_{k_{i}}$ models has a discrete symmetry group which is defined as follows
\ber
G_{tot}=\prod_{i=1}^{5}\mathbb{Z}_{k_{i}+2}=\lbrc \prod_{i=1}^{5} \hat{g}_{i}^{w_{i}}, \ w_{i}\in \mathbb{Z}, 
\hat{g}_{i}=\exp{(\im 2\pi J_{(i),0})}\rbrc.
\label{2.Gtot}
\enr

The admissible group is a subgroup of $G_{tot}$ which is defined as follows:
\ber
G_{adm}=\lbrc &\vw=\sum_{a=0}^{K-1}m_{a}\vbt_{a}, \ m_{a}\in\mathbb{Z},
\
\sum_{i}{\bet_{ai}\ov k_{i}+2}\in\mathbb{Z}, \ \bet_{ai}\in\mathbb{Z}\rbrc\subset G_{tot},
\nmb
&\vbt_{0} \equiv (1,1,1,...,1)\in G_{adm}. 
\label{2.Gadm}
\enr
Thus, the 5-dimensional vectors $\vbt_{a}$ are the generators of the admissible group, so that an arbitrary element $\vw\in G_{adm}$ can be decomposed in terms of these generators. The admissible group is defined in such a way to preserve a nowhere-vanishing holomorphic $(3,0)$-form on the CY hypersurface.

\vskip 10pt
\paragraph{Spectral flow construction of states in $N=(2,2)$ Minimal Model.}

In \cite{BBP} (see also \cite{BP}), the spectral flow have been used to construct the orbifold model primary fields. Here we recall this construction.

The spectral flow \cite{SS} is given by the one-parametric family of automorphisms
\ber
&\tilde{G}^{\pm}_{r}=U^{-t}G^{\pm}_{r}U^{t}=G^{\pm}_{r\pm t},\\
&\tilde{J}_{n}=U^{-t}J_{n}U^{t}=J_{n}+{c\ov3}t\dlt_{n,0},\\
&\tilde{L}_{n}=U^{-t} L_{n} U^{t}=L_{n}+tJ_{n}+{c\ov 6}t^{2}\dlt_{n,0},
t\in {1\ov 2}+\mathbb{Z}
\label{2.Sflow}
\enr
of the $N=2$ Virasoro superalgebra. 

For the minimal model $M_{k}$ spectral flow can be used to construct all primary fields \cite{FST}. Indeed, the state 
\ber
V_{l,t}=(UG^{-}_{-{1\ov 2}})^{t}\Phi^{c}_{l}, \quad 0\leq t\leq l.
\label{2.Prim1}
\enr
gives spectral flow realization of the primary state $\Phi^{NS}_{l,q}$, where $q=l-2t$.
For the purposes of the orbifold construction we need to extend this formula. 
Namely, we define the state
\ber
V_{l,t}=(UG^{-}_{-{1\ov 2}})^{t-l-1}U(UG^{-}_{-{1\ov 2}})^{l}\Phi^{c}_{l}, \quad l+1\leq t\leq k+1.
\label{2.Prim2}
\enr
It can be checked that  the state $V_{l,t}$ gives the spectral flow realization of the primary state $\Phi^{NS}_{\tld{l},\tld{q}}$, where $\tld{l}=k-l$, $\tld{q}=k+2+l-2t$. 

Notice the spectral flow the periodicity property \cite{FST}, (see also \cite{FeS})
\ber
U^{k+2}\approx 1.
\label{2.Period}
\enr 
In what follows we will use the notation $V_{l,t}$ instead of $\Phi^{NS}_{l,q}$ to emphasize that we are dealing with the spectral flow realization (\ref{2.Prim1}), (\ref{2.Prim2}) of primary states of Minimal model.

The spectral flow construction of the primary  states in the $R$-sector can be obtained by applying the operator $U^{1\ov 2}$ to the expressions (\ref{2.Prim1}), (\ref{2.Prim2}).

It is helpful now to refine the formulas (\ref{2.Prim1}), (\ref{2.Prim2}) to see when the spectral flow construction gives the chiral or anti-chiral primary states of minimal model. We have
\ber
&t=0:V_{l,0}(z)=\Phi^{c}_{l}(z)\ \text{is a chiral primary},
\nmb
&t=l:V_{l,l}(z)=(UG^{-}_{-{1\ov 2}})^{l}\Phi^{c}_{l}(z)\approx\Phi^{a}_{l}(z)\ \text{is an anti-chiral primary},
\nmb
&t=l+1:V_{l,l+1}(z)=U(UG^{-}_{-{1\ov 2}})^{l}\Phi^{c}_{l}(z)\approx\Phi^{c}_{\tld{l}}(z),\ \tld{l}=k-l,
\ \text{is a chiral primary},
\nmb
&t=k+1:V_{l,k-l}(z)=(UG^{-}_{-{1\ov 2}})^{k-l}U(UG^{-}_{-{1\ov 2}})^{l}\Phi^{c}_{l}(z)\approx\Phi^{a}_{\tld{l}}(z), 
\,\tld{l}=k-l, \ \text{is an anti-chiral primary}.
\nmb
\label{2.Mca}
\enr

Here it is worth noting that, obviously, instead of (\ref{2.Prim1}), (\ref{2.Prim2}) we could propose a construction that begins with antichiral primary fields. This mirror construction will be the key point of discussion in the next section.

\vskip 10pt
\paragraph{Spectral flow construction of states in the orbifold model.}

The spectral flow realizations (\ref{2.Prim1}),(\ref{2.Prim2}) are very convenient to build the primary fields in the orbifold model. The construction is given in three steps. 

In the first step, we use the elements $\vw$ of the admissible group (\ref{2.Gadm})
for expanding the state space of the product model $M_{\vk}$ by adding in $NS$ sector the twisted fields of the form
\ber
\Psi^{NS}_{\vl,\vt,\vw}(z,\bar{z})=V_{\vl,\vt+\vw}(z)\bar{V}_{\vl,\vt}(\bar{z}),
\label{2.Primo}
\enr
where
\ber
\bar{V}_{\vl,\vt}(\bar{z})=\prod_{i}\bar{V}_{l_{i},t_{i}}(\bar{z}),\
\bar{V}_{l_{i},t_{i}}(z)=(UG^{-}_{-{1\ov 2}})_{i}^{t_{i}}\bar{\Phi}^{c}_{l_{i}}(z), \ 0\leq t_{i}\leq l_{i},
\label{2.Primo1}
\enr
and
\ber
V_{\vl,\vt+\vw}(z)=\prod_{i=1}^{5}V_{l_{i},t_{i}+w_{i}}(z),
\enr
where
\ber
\begin{aligned}
&V_{l_{i},t_{i}+w_{i}}(z)=\begin{cases}(UG^{-}_{-{1\ov 2}})_{i}^{t_{i}+w_{i}}\Phi^{c}_{l_{i}}(z), 
\qquad \text{if}\quad \ 0\leq t_{i}+w_{i}\leq l_{i}, \\\\
(UG^{-}_{-{1\ov 2}})_{i}^{t_{i}+w_{i}-l_{i}-1}
U_{i}(UG^{-}_{-{1\ov 2}})_{i}^{l_{i}}\Phi^{c}_{l_{i}}(z), 
\quad \text{if}\quad \ l_{i}+1\leq t_{i}+w_{i}\leq k_{i}+1.
 \end{cases}
\label{2.Primo2}
\end{aligned}
\enr

In the second step, we require the mutual locality of the fields obtained above. It gives the equations
\ber
\sum_{i}{\bet_{ai}(q_{i}-w_{i})\ov k_{i}+2}\in\mathbb{Z}, \ a=0,...,K-1.
\label{2.NSNSloc}
\enr 
Thus, the field $\Psi^{NS}_{\vl,\vt,\vw}(z,\bar{z})$ will be mutually local with all other fields in $NS$ sector if (\ref{2.NSNSloc}) is fulfilled.
In other words, this system of equations singles out the admissible $(\vl,\vt)$ (or $(\vl,\vq)$) of $NS$ primary fields $\Psi^{NS}_{\vl,\vt,\vw}(z,\bar{z})$ which appear in the twisted sector $\vw$ of the orbifold model.

The third step of the construction is simple: we generate $R$ sector fields as
\ber
\Psi^{R}_{\vl,\vt,\vw}(z,\bar{z})=\prod_{i}U^{1\ov 2}_{i}\bar{U}^{1\ov 2}_{i}
\Psi^{NS}_{\vl,\vt,\vw}(z,\bar{z}).
\label{2.RPrimo}
\enr

These fields together with the mutually local $NS$ fields constructed before can be considered as a set of primary fields of the orbifolds in the sense that all the other fields of the orbifold are generated by applying to them the creation operators of the $N=(2,2)$ Virasoro superalgebras of $M_{k_{i}}$ models.

It is clear that $NS$ sector fields of the orbifold are mutually local (by construction), while $R$ sector fields are mutually quasi-local among themselves and with $NS$ sector fields. In other words, the quasi-locality structure of the orbifold is the same as for any $N=(1,1)$ superconformal model.

The orbifold model satisfy all the requirements of Conformal Bootstrap, as shown in \cite{BBP}. In particular, the OPE is closed on the set of $NS$ sector fields restricted by the locality equations (\ref{2.NSNSloc}) and this entails the OPE closure of the total set of orbifold model fields. 

\vskip 10pt
\paragraph{The chiral rings states.}

In this paragraph we recall the construction of chiral-chiral and anti-chiral-chiral states using the general expressions (\ref{2.Primo})-(\ref{2.Primo2}).

In order to get $(c,c)$ primary field in orbifold model one must have  $(c,c)$ primary field in each  $M_{k_{i}}$ factor of the composite model $M_{\vk}$. Therefore, according to the general formula (\ref{2.Primo}) we can see that in the anti-holomorphic sector the chiral primary state is given by the first line in (\ref{2.Mca}) in each factor $M_{k_{i}}$, while in the holomorphic sector the chiral primary state is given by the first line or by the third line in (\ref{2.Mca}). It gives the 
\vskip .3cm

\paragraph{Algorithm for finding $(c,c)$ fields in the twisted $\vw$ sector.}

{\bf 1.} For each $N=0,1,2,3$ to find all the vectors $\vl=(l_{1},...,l_{5})$ satisfying the equations
\ber
\sum_{i}{l_{i}\ov k_{i}+2}=N, \quad \sum_{i}\bet_{ai}{l_{i}-w_{i}\ov k_{i}+2}\in\mathbb{Z},\quad
a=0,...,K-1.
\label{2.CC1}
\enr
{\bf 2.} Among the found vectors $\vl$, we leave only those for which the set of numbers $w_i$ satisfies the following relations 
\ber
\begin{aligned} 
 w_{i}&= 0 \mod\ k_{i}+2, \ & \text{or} \ w_{i}&=l_{i}+1 \mod\ k_{i}+2.
\label{2.CC2}
\end{aligned}
\enr


Similarly, one can obtain the algorithm of finding $(a,c)$ primaries in the orbifold. In this case one should take $(a,c)$ primary field in each factor $M_{k_{i}}$ of the composite model $M_{\vk}$. Therefore, in the anti-holomorphic sector the chiral primary state is given again by the first line from (\ref{2.Mca}) in each factor $M_{k_{i}}$, while in the holomorphic sector anti-chiral primary state is given by the second line or by the fourth line from (\ref{2.Mca}). Hence, taking into account (\ref{2.Primo})-(\ref{2.Primo2}) we obtain the

\paragraph{Algorithm for finding $(a,c)$ fields in the twisted $\vw$ sector.}

{\bf 1.} For each $N=0,...,3$ find out all the vectors $\vl=(l_{1},...,l_{5})$ satisfying the equations
\ber
\sum_{i}{l_{i}\ov k_{i}+2}=N, \quad \sum_{i}\bet_{ai}{l_{i}-w_{i}\ov k_{i}+2}\in\mathbb{Z} ,\quad
a=0,...,K-1. 
\label{2.AC1}
\enr
{\bf 2.}  Among the found vectors $\vl$, we keep only those for which the set of numbers $w_i$ satisfies to the following relations
\ber
\begin{aligned}
 w_{i}&= l_{i} \mod\ k_{i}+2, \ &\text{or} \ w_{i}&=k_{i}+1 \mod\ k_{i}+2.
\label{2.AC2}
\end{aligned}
\enr

For more details see \cite{BBP}, where the $(c,c)$ and $(a,c)$ fields has been constructed for several examples of orbifold mirror pairs.

\vskip 10pt

\section{Mirror spectral flow construction and mirror orbifold.}
\label{sec:3}

In this section we move on to the main point of the paper. We present mirror spectral flow construction, which together with mutual locality requirement leads to the mirror orbifold model.

 Consider first the mirror version of formulas (\ref{2.Prim1}), (\ref{2.Prim2}), which starts with the anti-chiral state:
\ber
V_{l,t}=(U^{-1}G^{+}_{-{1\ov 2}})^{l-t}\Phi^{a}_{l}, \quad 0\leq t\leq l,
\label{3.Prim1}
\enr
\ber
V_{l,t}=(U^{-1}G^{+}_{-{1\ov 2}})^{k+1-t}U^{-1}(U^{-1}G^{+}_{-{1\ov 2}})^{l}\Phi^{a}_{l}, \quad l+1\leq t\leq k+1.
\label{3.Prim2}
\enr
Their validity is coming from the analysis of singular vectors in the minimal $N=2$ Virasoro superalgebra representations \cite{FST}. Another way to get these expressions is to invert the expressions from the second and fourth lines of (\ref{2.Mca}). 
 
The mirror spectral flow construction of fields in the orbifold model starts with anti-chiral primary state in holomorphic factor of (\ref{2.Primo}) and uses (\ref{3.Prim1}), (\ref{3.Prim2}), while the anti-holomorphic factor remains intact. Thus, for the anti-holomorphic factor of $\Psi^{NS}_{\vl,\vt,\vw}(z,\bar{z})$ we still have the expression (\ref{2.Primo1}), while for the holomorphic factor we apply (\ref{3.Prim1}) or (\ref{3.Prim2}) to get:
\ber
\begin{aligned}
&V_{\vl,\vt+\vw}(z)=\prod_{i=1}^{5}V_{l_{i},t_{i}+w_{i}}(z),
\nmb
&V_{l_{i},t_{i}+w_{i}}(z)=
\begin{cases}
(U^{-1}G^{+}_{-{1\ov 2}})_{i}^{l_{i}-t_{i}-w_{i}}\Phi^{a}_{l_{i}}(z), 
\qquad \text{if}\quad \ 0\leq t_{i}+w_{i}\leq l_{i},
\\\\
(U^{-1}G^{+}_{-{1\ov 2}})_{i}^{k_{i}+1-t_{i}-w_{i}}
U_{i}^{-1}(U^{-1}G^{+}_{-{1\ov 2}})_{i}^{l_{i}}\Phi^{a}_{l_{i}}(z), 
\quad \text{if}\quad \ l_{i}+1\leq t_{i}+w_{i}\leq k_{i}+1.
\end{cases}
\end{aligned}
\enr  


Making the involution
\ber
G^{\pm}(z)\rightarrow G^{\mp}(z), \ J(z)\rightarrow -J(z), \ U(z)\rightarrow U^{-1}(z),
\ T(z)\rightarrow T(z).
\label{3.Inv}
\enr
we transform the expressions above back into the old form but with different twists
\ber
V_{\vl,\vt+\vw^{*}}(z)=\prod_{i=1}^{5}V_{l_{i},t_{i}+w^{*}_{i}}(z),
\label{3.Mirror1}
\enr
where
\ber
\begin{aligned}
&V_{l_{i},t_{i}+w^{*}_{i}}(z)=
\begin{cases}
(UG^{-}_{-{1\ov 2}})_{i}^{t_{i}+w^{*}_{i}}\Phi^{c}_{l_{i}}(z),\quad w^{*}_{i}=l_{i}-2t_{i}-w_{i},
\\\\
(UG^{-}_{-{1\ov 2}})_{i}^{t_{i}+w^{*}_{i}-l_{i}-1}
U_{i}(UG^{-}_{-{1\ov 2}})_{i}^{l_{i}}\Phi^{c}_{l_{i}}(z), 
\quad w^{*}_{i}=k_{i}+2+l_{i}-2t_{i}-w_{i}\approx l_{i}-2t_{i}-w_{i}
\end{cases}
\label{3.Mirror2}
\end{aligned}
\enr 
(note that $w^{*}_{i}$ is $mod \ k_{i}+2$ defined). Recovering the anti-holomorphic factors we obtain another representation for the $NS$ sector fields of the orbifold model:
\ber
\Psi^{NS}_{\vl,\vt,\vw^{*}}(z,\bar{z})=V_{\vl,\vt+\vw^{*}}(z)\bar{V}_{\vl,\vt}(\bar{z}).
\label{3.Primdual}
\enr
The $R$ sector primary fields now take the form
\ber
\Psi^{R}_{\vl,\vt,\vw^{*}}(z,\bar{z})=\prod_{i}U^{-{1\ov 2}}_{i}\bar{U}^{1\ov 2}_{i}
\Psi^{NS}_{\vl,\vt,\vw^{*}}(z,\bar{z}).
\label{3.RPrimo}
\enr
Thus, the mirror spectral flow construction (together with the involution (\ref{3.Inv})) gives another representation for the fields of the orbifold model, where the twisted sectors are numbered by another set of twist vectors $w^{*}$. In other words, the mirror spectral flow construction looks like the orbifold of $M_{\vk}$ which is made by using another group. More precisely, we have the following

\vskip 10pt
\leftline{\bf Proposition.}

For each $N=(2,2)$ superconformal orbifold model $M_{\vk}/G_{adm}$, the mirror spectral flow construction (\ref{3.Mirror1})-(\ref{3.Primdual}) defines new admissible group $G^{*}_{adm}$ and $N=(2,2)$ superconformal orbifold model $M_{\vk}/G^{*}_{adm}$ such that:

{\bf 1.} the group $G^{*}_{adm}$ is Berglund-Hubsh-Krawitz dual to $G_{adm}$;

{\bf 2.} the model $M_{\vk}/G^{*}_{adm}$ is isomorphic to the model $M_{\vk}/G_{adm}$;

{\bf 3.} $(c,c)$ ring ($(a,c)$ ring) of $M_{\vk}/G^{*}_{adm}$ coincides with $(a,c)$ ring ($(c,c)$ ring) of $M_{\vk}/G_{adm}$.

\vskip 10pt
\leftline{\bf Proof.}
 
{\bf 1,2.} For an arbitrary vector $\vw^{*}$ and arbitrary $\vw\in G_{adm}$ we have from (\ref{3.Mirror2})
\ber
\sum_{i}{w_{i}w^{*}_{i}\ov k_{i}+2}=
\begin{cases}
\sum_{a}w^{a}\sum_{i}{\bet_{ai}(l_{i}-2t_{i}-w_{i})\ov k_{i}+2},\quad \text{if} \ w^{*}_{i}=l_{i}-2t_{i}-w_{i},
\\\\
\sum_{a}w^{a}\sum_{i}{\bet_{ai}(k_{i}+2+l_{i}-2t_{i}-w_{i})\ov k_{i}+2},
\quad \text{if} \ w^{*}_{i}=k_{i}+2+l_{i}-2t_{i}-w_{i}.
\end{cases}
\nmb
\enr
Thus, taking into account (\ref{2.NSNSloc}) we obtain another form of mutual locality equations
for the original orbifold $M_{\vk}/G_{adm}$:
\ber
\sum_{i}{w_{i}w^{*}_{i}\ov k_{i}+2}\in\mathbb{Z}, \ \vw\in G_{adm}.
\label{3.Gdual}
\enr
As a particular case of (\ref{3.Gdual}) we obtain
\ber
\sum_{i}{w^{*}_{i}\ov k_{i}+2}\in\mathbb{Z}.
\label{3.Gadm}
\enr
This equation means that the new twisted vectors $\vw^{*}=\vl-2\vt-\vw$ satisfy admissible group equations (\ref{2.Gadm}). Moreover, the vector $(1,1,1,1,1)$ arises among $\vw^{*}$. Indeed, if $\vl=\vt=0$, $w=(k_{1}+1,...,k_{5}+1)$, the twist vector $\vw^{*}=-\vw=-(k_{1}+1,...,k_{5}+1)\approx (1,1,1,1,1)$ and the corresponding state fulfills mutual locality equations (\ref{2.NSNSloc})
\ber
\sum_{i}\bet_{ai}{-k_{i}-1\ov k_{i}+2}=\sum_{i}\bet_{ai}(-1+{1\ov k_{i}+2})\in\mathbb{Z}
\nmb
\enr
(where the admissible group equation for $\vbt_{a}$ has been used.)

It is easy to see that the new twists form an abelian group because for any pair $\vw^{*}_{1}$ and $\vw^{*}_{2}$, the vector $\vw^{*}_{1}+\vw^{*}_{2}$ arises among the set of $\vw^{*}$. It indeed follows because 
\ber
\vw^{*}_{1,2}=\vl_{1,2}-2\vt_{1,2}-\vw_{1,2} =\vq_{1,2}-\vw_{1,2},
\nmb
\enr
but $\vq$ is conserved while $\vw$ is additive when we consider the OPE of the corresponding fields. Hence, the twists $\vw^{*}$ are additive w.r.t. the OPE (in other words, the mutual locality conditions (\ref{2.NSNSloc}) are consistent with the OPE). Thus, we have shown that the set of new twists $\vw^{*}$, arising in (\ref{3.Mirror2}) form an admissible group, which we denote as $G^{*}_{adm}$. 
Hence, the mirror spectral flow construction (\ref{3.Primdual}), (\ref{3.Inv}) defines new orbifold model $M_{\vk}/G^{*}_{adm}$, because (\ref{3.Gdual}) can also be considered as mutual locality equations for this new model. By the construction, this new $N=(2,2)$ superconformal model is isomorphic to the original one.

The group $G^{*}_{adm}$ is obviously the maximal dual group to $G_{adm}$. 

Finally note that mutual locality equations (\ref{3.Gdual}) are nothing else but the definition of dual group of Berglund, Hubsh and Krawitz \cite{Kra}.

{\bf 3.} Let us show that equations from $(c,c)$ algorithm for dual orbifold model $M_{\vk}/G^{*}_{adm}$ are equivalent to the equations from $(a,c)$ algorithm for the original model. 

Because of $t_{i}$ must be zero for the $(c,c)$ field of the dual model, the conditions
(\ref{2.CC2}) for the dual twists take the form
\ber
w^{*}_{i}= 
0 \Leftrightarrow w_{i}=l_{i}, 
\nmb
\text{or}
\ l_{i}+1=w^{*}_{i}\Leftrightarrow w_{i}=k_{i}+1.
\label{4.CC}
\enr
where we have used expressions for $w^{*}_{i}$ from (\ref{3.Mirror2}).
Thus, $(c,c)$ equations for the dual model are nothing else but $(a,c)$ equations (\ref{2.AC2}) for the original orbifold.

Let us now consider $(a,c)$ algorithm equations for the dual model. Again, $t_{i}=0$ so that (\ref{2.AC2}) equations for the dual twists take the form
\ber
w^{*}_{i}=l_{i}\Leftrightarrow w_{i}=0,
\nmb
\text{or} \
w^{*}_{i}=k_{i}+1\Leftrightarrow w_{i}=l_{i}+1,
\label{4.AC}
\enr
where we have used again the expressins for $w^{*}_{i}$ from (\ref{3.Mirror2}) and have taken into account that $w_{i}$ is $mod \ k_{i}+2$ defined.
Hence, $(a,c)$ equations for dual model are nothing else but $(c,c)$ equations (\ref{2.CC2}) for the original orbifold. It finishes the proof of Proposition.

The {\bf Proposition} thus gives the explicit construction of mirror pair of CY orbifolds 
$(M_{\vk}/G_{adm}$, $M_{\vk}/G^{*}_{adm})$, which correspond to the same $N=(2,2)$ superconformal model. Moreover, the expressions (\ref{2.Primo})-(\ref{2.Primo2}) and (\ref{3.Mirror1})-(\ref{3.Primdual}) provide the explicit mapping between the states of this pair, such that $(c,c)$ ($(a,c)$) ring of $M_{\vk}/G_{adm}$ orbifold is mapped to $(a,c)$ ($(c,c)$) ring of $M_{\vk}/G^{*}_{adm}$ orbifold.

While the spectral flow construction and its mirror have been based only at one point in the respective moduli
space (namely, the point which has an interpretation as an orbifold of composite minimal models),
the result necessarily extends to all other nearest points of moduli space. 

Notice also that the spectral flow construction of mirror pairs is not limited to the product of 5 models and can be extended for the composite models of a more general type. 

The spectral flow construction of mirror pairs of orbifolds is very close to the algorithm which was discussed in \cite{GrPl}. Indeed, it is easy to see that the additional $\mathbb{Z}_{k_{i}+2}$ orbifolding for each $M_{k_{i}}$  (see \cite{GrPl}) is equivalent to application of mirror spectral flow construction, instead of the original spectral flow construction from section 2. Although the mirror spectral flow construction can be viewed as the result of $\mathbb{Z}_{k_{i}+2}$ orbifoldings, we prefer not to do all these intermediate steps, because, in our opinion, the mirror spectral flow construction looks more clear and natural in the context of the approach advocated in \cite{BBP}, \cite{BP}. 
Besides, the orbifolding from \cite{GrPl} does not work directly for non Fermat-type polynomials \cite{BH}, while the construction of mirror pairs discussed above, is, as we hope, more general.



\section*{Acknowledgments} 

Author acknowledges Prof. A. A. Belavin for helpful discussions. 
The work was carried out at Landau Institute for Theoretical Physics in the framework of the state assignment No. 0029-2019-0004.


\end{document}